\documentclass[twocolumn,showpacs,preprintnumbers,amsmath,amssymb]{revtex4}

\usepackage{graphicx}
\usepackage{dcolumn}
\usepackage{bm}

\begin{document}

\preprint{}

\title{A  scheme for dense coding in the non-symmetric quantum channel}

\author{Fengli Yan $^{1,2}$ and Meiyu Wang $^2$}
 \affiliation{%
$^1$ CCAST (World Laboratory), P.O. Box 8730, Beijing 100080,
China\\
$^2$ Department of Physics, Hebei Normal University, Shijiazhuang
050016, China
}%

\date{\today}

\begin{abstract}
We investigate the dense coding in the case of  non-symmetric  Hilbert spaces of the sender and receiver's
particles sharing the  quantum maximally entangled state. The efficiency of classical information gain is also
considered. We conclude that when a more level particle is with the sender, she can get a non-symmetric quantum
channel from a symmetric one by entanglement transfer. Thus the efficiency of information transmission is
improved.
\end{abstract}

\pacs{03.67.-a, 89.70.+c}
\maketitle

\section{Introduction  }
The  quantum entanglement state [1, 2] is an important tool distinguishing the quantum mechanics from the
classical physics. Due to the non-local properties of quantum entanglement state, it has been widely used in
many fields such as quantum teleportation \cite {s3}, quantum dense coding \cite {s4} and quantum key
distribution \cite {s5}. Here we will focus our attention on quantum dense coding. In 1992, Bennett and Wiesner
proposed the first scheme for the quantum dense coding \cite {s4}, in which  Bell
 states are used as quantum channel. In 1996, quantum dense coding was
experimentally presented by Mattle $et$  $al$  in an optical system \cite {s6}. Recently Liu $et$ $al$ presented
a protocol for dense coding with multi-level entangled states \cite {s7}. However in these  schemes, the quantum
channels are  symmetric, that is to say, the dimension of the Hilbert space of  the  particle with sender is the
same as that of the particle with receiver. In 1998, Bose $et$ $al$ generalized Bennett and Wiesner's scheme for
dense coding into multi-parties, one receiver intends to receive messages from $N$ senders \cite {s8} . In this
case, the quantum channel is non-symmetric. The particles with senders are in $2^N$- dimensional Hilbert space,
while the particle with 2-dimensional Hilbert space belongs to receiver.  In present paper, we give a scheme for
quantum dense coding in the case of  non-symmetric Hilbert spaces  of the sender and receiver's particles
sharing the  quantum maximally entangled state, and consider the efficiency of classical information gain.

\section{A  scheme for dense coding in the non-symmetric quantum channel}

   In Bennett and Wiesner's first scheme for the quantum dense
   coding, sender Alice and receiver Bob share a pair of
   entangled particles in the Bell state. Alice performs one of
   the four 1-qubit unitary operations given by the identity $I$
   or the Pauli matrices $(\sigma_x, i\sigma_y, \sigma_z)$ on her
   particle. Each of the unitary operations maps the Bell state to
   a different member of the four Bell states. Then Alice sends
   her particle to Bob. Bob can obtain two bits of classical
   information from the joint measurement on two particles. Now
   we would like to generate the Bennett and Wiesner's symmetric scheme to the
   non-symmetric one.

   To present our scheme clearly, let us first begin with dense
   coding between two parties in  $3\times 2$-dimension. Suppose
   the particle 1 in 3-dimensional Hilbert space belongs to Alice, and the
   particle 2 in 2-dimensional Hilbert space is with  Bob. They share the maximally
   entangled state:
   \begin{equation}\label{}
    |\Psi_{00}\rangle_{12}=\frac {1}{\sqrt 2}(|00\rangle+|11\rangle)_{12}.
\end{equation}
Through simple calculation, it can be shown that the single-body operators on particle 1:
\begin{equation}\label{}
\begin{array}{l}
 U_{00}=\left [ \begin{array}{ccc} 1&0&0\\
0&1&0\\
0&0&1\\\end{array} \right ], ~~~~U_{01}=\left [ \begin{array}{ccc} 1&0&0\\
0&-1&0\\
0&0&1\\\end{array} \right ], \\[0.5cm]
U_{10}=\left [ \begin{array}{ccc} 0&0&1\\
1&0&0\\
0&1&0\\\end{array} \right ],
~~~~U_{11}=\left [ \begin{array}{ccc} 0&0&1\\
1&0&0\\
0&-1&0\\\end{array} \right ],\\[0.5cm] U_{20}=\left [ \begin{array}{ccc} 0&1&0\\
0&0&1\\
1&0&0\\\end{array} \right ], ~~~~U_{21}=\left [ \begin{array}{ccc} 0&-1&0\\
0&0&1\\
1&0&0\\\end{array} \right ]
\end{array}
\end{equation}
will transform $|\Psi_{00}\rangle $ into the corresponding state
respectively:
\begin{equation}
   U_{00}|\Psi_{00}\rangle=\frac {1}{\sqrt 2}(|00\rangle+|11\rangle)=|\Psi_{00}\rangle,
\end{equation}
\begin{equation}
   U_{01}|\Psi_{00}\rangle=\frac {1}{\sqrt 2}(|00\rangle-|11\rangle)=|\Psi_{01}\rangle,
\end{equation}
\begin{equation}
   U_{10}|\Psi_{00}\rangle=\frac {1}{\sqrt 2}(|10\rangle+|21\rangle)=|\Psi_{10}\rangle,
\end{equation}
\begin{equation}
   U_{11}|\Psi_{00}\rangle=\frac {1}{\sqrt 2}(|10\rangle-|21\rangle)=|\Psi_{11}\rangle,
\end{equation}
\begin{equation}
   U_{20}|\Psi_{00}\rangle=\frac {1}{\sqrt 2}(|20\rangle+|01\rangle)=|\Psi_{20}\rangle,
\end{equation}
\begin{equation}
   U_{21}|\Psi_{00}\rangle=\frac {1}{\sqrt 2}(|20\rangle-|01\rangle)=|\Psi_{21}\rangle.
\end{equation}

Alice operates one of the above unitary transformations, and sends her particle to Bob. Bob takes only one
measurement in the base $\{|\Psi_{00}\rangle, |\Psi_{01}\rangle,\cdots, |\Psi_{21}\rangle\}$, and he will know
what operation Alice has done, or to say , what the messages Alice has encoded in the quantum state. As a
result, Bob gets $\log_26$ bits of information through only one measurement. Thus the dense coding is realized.

It is straightforward to generalize the above protocol to arbitrarily different dimension for two parties.
Assume that the Hilbert space of Alice's particle 1 is non-symmetric with that of Bob's particle 2, i.e. the
dimension $p$ of the Hilbert space of  particle 1 is not the same as the dimension $q$ of  particle 2.
 Without loss of generality, we choose $p>q$.  The non-symmetric quantum channel shared by Alice and Bob is the maximally entangled state:
\begin{equation}
  |\Psi_{00}\rangle_{12}=\frac {1}{\sqrt q}(|00\rangle+|11\rangle+\cdots+|q-1q-1\rangle)_{12}.
\end{equation}

Clearly, an orthogonal  base can be composed of the general Bell states  of particle 1 and 2
\begin{equation}|\Psi_{mn}\rangle=\sum_j {\rm
e}^{2\pi{\rm i}jn/q}|(j\oplus m){\rm mod} p\rangle_1\otimes|j\rangle_2/{\sqrt q},
\end{equation}
  where $m=0,1,\cdots, p-1;$ $n,j=0,1,\cdots, q-1$. The single-body operators on  particle 1
satisfying $U_{mn}|\Psi_{00}\rangle=|\Psi_{mn}\rangle$ may be explicitly written out:
\begin{equation} U_{mn}=\sum_j {\rm e}^{2\pi{\rm
i}jn/q}|(j\oplus m){\rm mod} p\rangle\langle j|,
\end{equation}
where $m=0,1,\cdots, p-1;$ $n,j=0,1,\cdots, q-1$. The procedure for realizing dense coding in this case is
similar to the above one. In order to send information, Alice performs one of the operations in (11) to her
particle and sends it to Bob. Bob then performs a collective measurement in the general Bell base $\{
|\Psi_{mn}\rangle_{12}|m=0,1,\cdots,p-1; q=0,1,\cdots,q-1\}$
  to find out what
Alice has done to particle 1 and therefore reads out the encoded message. In this case, Bob can gain
$\log_2p\times q$ bits  of information. If $p=q$, the quantum channel becomes symmetric, correspondingly the
bits of information become $\log_2p^2=2\log_2p$.

\section{improving the efficiency of the dense coding by
entanglement transfer}

When the quantum channel between Alice and Bob is symmetric, besides a more level particle is with Alice, then
the quantum channel may be transformed into non-symmetric one, and hence improve the efficiency of the dense
coding. For the sake of clearness, we start our discussion  from the following simple case.

Suppose the initial quantum channel between Alice and Bob is
\begin{equation}
 |\Phi\rangle_{12}=\frac {1}{\sqrt 2}(|00\rangle+|11\rangle)_{12},
\end{equation}
the particle 1 belongs to Alice, and the particle 2 to Bob, they are both in 2-dimensional Hilbert space.
Besides the particle 3 belonging to Alice is in 3-dimensional   Hilbert space, its initial state is
$|0\rangle_3$. So the initial state of the three particles is
\begin{equation}
 |\Psi\rangle_{123}=\frac {1}{\sqrt 2}(|00\rangle+|11\rangle)_{12}|0\rangle_3.
\end{equation}
First Alice applies the following unitary operation on her particles 1 and 3 under the base
$\{|00\rangle_{13},|01\rangle_{13},|02\rangle_{13},|10\rangle_{13},|11\rangle_{13},|12\rangle_{13}\}$:
\begin{equation}
  U_{13}=\left [\begin{array}{cccccc}
  1&0&0&0&0&0\\  0&1&0&0&0&0\\0&0&1&0&0&0\\0&0&0&0&1&0\\0&0&0&1&0&0\\0&0&0&0&0&1\\\end{array}
  \right ],
\end{equation}
correspondingly $|\Psi\rangle_{123}$ becomes
\begin{equation}
U_{13}|\Psi\rangle_{123}=\frac {1}{\sqrt
2}(|00\rangle_{13}|0\rangle_2+|11\rangle_{13}|1\rangle_2).
\end{equation}
Second Alice applies the following unitary operation on her particles 1 and 3:
\begin{equation}
  U'_{13}=\left [\begin{array}{cccccc}
  1&0&0&0&0&0\\  0&0&0&0&1&0\\0&0&1&0&0&0\\0&0&0&1&0&0\\0&1&0&0&0&0\\0&0&0&0&0&1\\\end{array}
  \right ],
\end{equation}
and hence equation (15) becomes
\begin{equation}
\begin{array}{l}
~~~~U'_{13}\frac {1}{\sqrt
2}(|00\rangle_{13}|0\rangle_2+|11\rangle_{13}|1\rangle_2)\\
=\frac
{1}{\sqrt
2}(|00\rangle_{23}+|11\rangle_{23})|0\rangle_1\\
=|0\rangle_1|\Phi'\rangle_{23}.\\\end{array}
\end{equation}
As a result, the entanglement between the particles 1 and 2 is transferred to  the entanglement between the
particles 2 and 3. Before this, using the quantum channel between the particles 1 and 2 for dense coding, Bob
gets $\log_2 4$ bits  of information, but after that, Bob can gain $\log_2 6$ bits of information, so the
efficiency of information transmission is increased.

In the following section, we consider the general case. Assume that Alice and Bob share the maximally entangled
state:
\begin{equation}
  |\Phi_{00}\rangle_{12}=\frac {1}{\sqrt q}(|00\rangle+|11\rangle+\cdots+|q-1q-1\rangle)_{12},
\end{equation}
 where the particle 1 is with Alice, and the particle 2 with Bob, they
are both in $q$-dimensional Hilbert space. Besides the particle 3 belonging to Alice is in $p$-dimensional
Hilbert space, its initial state is $|0\rangle_3$, and  $p>q$. Therefore  three particles will in the state
$|\Phi_{00}\rangle_{12}\otimes|0\rangle_3$. Alice performs the following unitary operations on her particles 1
and 3:
\begin{equation}
 U_{13}=\sum_{i=0}^{q-1}|ii\rangle\langle
i0|+\sum_{j=1}^{q-1}|j0\rangle\langle jj|+\sum_{m,n}|mn\rangle\langle mn|,
\end{equation}
where $m =0,1,\cdots,q-1$;  $n=1,2,\cdots,p-1,$ and $m\neq n$;
\begin{equation}
 U'_{13}=\sum_{i=0}^{q-1}|0i\rangle\langle
ii|+\sum_{j=1}^{q-1}|jj\rangle\langle 0j|+\sum_{m,n}|mn\rangle\langle mn|,
\end{equation}
in the sum of above equation, when $m =0,$ we let $n=q, q+1, \cdots, p-1$; when $m=1,2,\cdots,q-1$, then $n$
should take $n=0,1,\cdots, p-1$ with the condition $m\neq n$. After that
$|\Phi_{00}\rangle_{12}\otimes|0\rangle_3$ correspondingly becomes
\begin{equation}
 \begin{array}{l}
   ~~~~U'_{13}\otimes U_{13}|\Phi_{00}\rangle_{12}\otimes|0\rangle_3\\
   =
   \frac {1}{\sqrt q}(|00\rangle+|11\rangle+\cdots+|q-1q-1\rangle)_{23}\otimes|0\rangle_1.
\end{array}
\end{equation}
So  the entanglement transfer between the particles 1 and 2 and the particles 2 and 3 has been successfully
realized. In other words, a  symmetric quantum channel has been transformed   into the non-symmetric one.
Clearly, before this, Bob gets $\log_2 q^2$ bits of information, now Bob can gain $\log_2 pq$ bits of
information, and hence the efficiency of information transmission can be increased.

\section{conclusion and discussion}

  In summary, we have given a  scheme  to transform the
  symmetric level quantum maximally entangled state between two
  parties into non-symmetric one, and using it as the quantum
  channel for dense coding. We have also considered the efficiency
  of information transmission. Comparing  our scheme with one for
  symmetric dense coding, obviously ours  is more efficient. We hope this scheme can be realized by the
  experiment.

\begin{acknowledgements}
 This work was supported  by  Hebei Natural Science
Foundation under Grant No. 101094.
\end{acknowledgements}


\begin{thebibliography}{s2}
\bibitem{s1} A. Einstein, B. Podolsky, and N. Rosen, Phys. Rev. {\bf 47},  777 (1935).
\bibitem{s2} E. Schr{\"{o}}dinger, Proc. Cambridge Philos. Soc. {\bf 31}, 555 (1935).
\bibitem{s3} C.H. Bennett, {\it et al},  Phys. Rev. Lett. {\bf 70},
1895 (1993).
\bibitem{s4} C.H. Bennett and S.J. Wiesner,  Phys. Rev. Lett. {\bf 69}, 2881 (1992).
\bibitem{s5} A.K. Ekert,   Phys. Rev. Lett. {\bf 67}, 661 (1991).
\bibitem{s6} K. Mattle {\it et al,}  Phys. Rev. Lett. {\bf 76}, 4656 (1996).
\bibitem{s7} X.S. Liu {\it et al},  Phys. Rev.  {\bf A 65}, 022304  (2002).
\bibitem{s8} S. Bose, V. Vedral, and P.L. Knight,  Phys. Rev.  {\bf A 57}, 822 (1998).
\end{thebibliography}
\end{document}